\documentstyle[twocolumn,aps,prl,epsf]{revtex}
\newcommand{\figwidth}{3.375in}
\begin{document}
\draft
\twocolumn[\hsize\textwidth\columnwidth\hsize\csname@twocolumnfalse\endcsname

\title{
How the Replica-Symmetry-Breaking Transition Looks Like 
in Finite-Size Simulations}
\author{Koji Hukushima$^{1}$ and Hikaru Kawamura$^{2}$}
\address{
$^1$Institute for Solid State Physics, University of Tokyo, 
5-1-5 Kashiwa-no-ha, Kashiwa, Chiba 277-8581, JAPAN\\ 
$^2$  Department of Earth and Space Science, Faculty of Science, 
Osaka University, Toyonaka, Osaka 560-0043, ~JAPAN}
\date{\today}
\maketitle
\begin{abstract}
Finite-size effects in the mean-field Ising spin glass and the mean-field
three-state Potts 
 glass are investigated   by Monte Carlo simulations.
In the thermodynamic limit, each model is known to exhibit 
a continuous phase transition into
the ordered state with a full and a one-step replica-symmetry breaking (RSB),
respectively. In the Ising case, 
Binder parameter $g$ calculated for various
finite sizes remains positive 
at any temperature and
crosses at the 
transition point, while in the Potts
case $g$ develops a negative dip without showing a crossing in the $g>0$ 
region.
 By contrast, 
non-self averaging  parameters always remain positive
and show a clear crossing at the transition
 temperature in both cases. 
Our finding suggests that care should be taken in interpreting the
numerical data of the Binder parameter, 
particularly when  the system exhibits a
one-step-like RSB.
\end{abstract}
\vskip2pc
]

\section{Introduction}
The concept of replica-symmetry breaking (RSB)\cite{ReviewsRSB} gives us
new insight into the character of the ordered state of
complex systems such as  spin glasses (SG)\cite{ReviewsSG} and real
structural glasses\cite{ReviewsG}.
Systems exhibiting the RSB  can roughly be divided into two categories 
depending on their breaking patterns: One is  a full or hierarchical RSB and
the other is a one-step RSB.
In both cases, there are many different equilibrium states unrelated by
global symmetry of the Hamiltonian, and an
overlap between these states plays an important role 
in describing
the ordered state.

In the case of one-step RSB\cite{ReviewsRSB}, the overlap $q$ takes
only two values in the thermodynamic limit, namely, 
either a self-overlap equal to the 
Edwards-Anderson
order parameter, $q=q_{{\rm EA}}$, or a non-self-overlap  usually equal
to zero, $q=0$. The overlap distribution function $P(q)$ consists of 
two distinct
delta-function peaks, one at $q=q_{{\rm EA}}$ and the other at $q=0$.
One-step RSB transitions could be either continuous or first-order, 
either with  or without a finite discontinuity in 
$q_{{\rm EA}}$ at the transition.
Examples of the first-order one-step RSB 
transition may be the mean-field $p$-spin glass
with $p>2$, the random energy model, and the mean-field $p$-state Potts
glass with $p>4$, while those of the continuous one-step RSB 
transition may be the
mean-field $p$-state Potts
glass with $2.8<p\leq 4$.

In the case of the full RSB, by contrast, 
possible values of the overlap are
distributed continuously in a certain range, and the states are organized in
a hierarchical manner. The overlap distribution function
has a continuous
plateau at $q< q_{{\rm EA}}$ in addition to the delta-function peak at
$q=q_{{\rm EA}}$.
Well-known example of this category is 
the standard mean-field Ising SG, namely,
the Sherrington-Kirkpatrick (SK) model. 
In some special cases, the admixture of the above twos, where
the overlap distribution function has a continuous plateau together with
the delta-function peak at $q=0$ (and the one at $q=q_{{\rm EA}}$), is
also possible. An example of this may be the mean-field $p$-state Potts glass
with $2<p<2.8$.

Recent interest in SG studies has been focused largely on the validity of 
applying the RSB idea  established in some mean-field models to more 
realistic short-range SG models. 
In almost all such studies, the three-dimensional (3D) Ising SG model
has been employed.
Indeed, some researchers have claimed that 
the full or hierarchical RSB as observed in the SK model\cite{Parisi} 
is also realized in
realistic 3D SG\cite{RSB3d}, while other researchers
have claimed, based on an alternative droplet picture\cite{FH}, 
that the  ordered state of realistic 3D SG is unique up to global symmetry 
of the Hamiltonian, without showing  RSB
of any kind\cite{Newman,Moore}. Thus, intensive debate has 
continued between these two scenarios 
as to the true nature of the SG ordered state of 
3D short-range systems.

Meanwhile, the one-step RSB has been discussed mainly
with interest in its close connection to
structural glasses rather than SG magnets\cite{Kirkpatrick,Parisi2}. 
Recently, however,   one-step RSB features have been found 
unexpectedly by the present
authors in the chiral-glass state of a 3D Heisenberg SG\cite{HK}. 
According to the chirality mechanism of experimental SG
transitions
based on the spin-chirality decoupling-recoupling scenario
\cite{CG},  
the SG ordered state and the SG 
phase transition of real Heisenberg-like SG magnets possessing 
weak but nonzero magnetic anisotropy 
are governed by the chirality ordering of the fully isotropic system
which is ``revealed'' by the weak magnetic anisotropy, 
{\it not by the spin ordering\/} which has been ``separated'' 
in the fully isotropic case from the chirality ordering. 
Then, the observation of Ref.~\cite{HK}
means that the SG ordered state of most 
of real SG magnets should also exhibit such one-step RSB-like features. Note 
that such a picture of the SG ordered state  
contrasts with  the standard pictures discussed so far,
either the droplet picture without the RSB
or the SK picture with the full RSB.

Under such circumstances, further studies of the nature of
the possible RSB in 3D short-range SG models are clearly required.
Since we are usually forced to employ numerical simulations to investigate
3D short-range models, and since
numerical simulations are often hampered by severe finite-size effects,
we feel it worthwhile to further clarify by numerical simulations
the finite-size effects in some
{\it mean-field models\/} which are exactly known to exhibit  RSB transitions
in the thermodynamic limit. In particular, the question how the
one-step and full RSB transitions look like 
in finite-size simulations is
of both fundamental and practical interest. 
Such information would be of much help as a reference
in interpreting the numerical data obtained for finite-dimensional
short-range  SG models. 

In the present paper, we choose two mean-field SG models
exactly known to exhibit a {\it continuous\/} (second-order)
phase transition in
the thermodynamic limit: One is the SK model which shows the full RSB,
and the other is the mean-field three-state ($p=3$) Potts-glass model which
shows the one-step RSB. We calculate by Monte Carlo simulations
several quantities which have widely been 
used in identifying the phase transition, 
including the spin-glass
order parameter and the Binder parameter, together with the
quantities recently introduced to represent the non-self-averaging
character of the ordered state. By carefully examining 
the size dependence of these quantities,
comparison is made between the two types of RSB. 
Our results have revealed that the Binder parameter of the one-step RSB system
shows the behavior very different from the standard behavior, giving 
warning about the interpretation of the numerical data for relevant short-range
systems.

\section{models}
The  mean-field $p$-state Potts-glass model is defined by the Hamiltonian, 
\begin{equation}
  \label{eqn:model}
  {\cal H} = -p\sum_{i<j}^{N}J_{ij}\delta_{n_i,n_j},
\end{equation}
where $n_i$ denotes a Potts-spin variable 
at the $i$-th site which takes $p$ distinct states,
and $N$ is the total number of Potts spins. 
The exchange interaction $J_{ij}$ is an
independent random Gaussian variable with zero mean and variance
$J^2/N$.
The model with $p=2$ is equivalent to the  SK model.
In the present study, we focus our attention on the standard SK model 
corresponding to $p=2$
and the three-state Potts-glass model corresponding to $p=3$.
Although the thermodynamic properties of an infinite system
have been rather well understood by the calculation based on a replica 
technique\cite{ReviewsRSB,Parisi,Elderfield,Gross},  its finite-size
properties have been much less understood.

It is convenient to use an equivalent simplex spin representation where
the Potts spin $n_i$ is written in terms of a $p-1$ dimensional unit vector
$\vec{S_i}$, 
which satisfies 
$\vec{S_i}\cdot\vec{S_j}=\frac{p\delta_{n_in_j}-1}{p-1}$, 
\begin{equation}
  \label{eqn:simplex}
  {\cal H} = -(p-1)\sum_{i<j}^{N}J_{ij}{\vec S_i}\cdot {\vec S_j}.
\end{equation}
In the particular
case of $p=2$, $\vec{S_i}$ simply reduces to the  
one-component Ising variable $S_i=\pm 1$, 
and the Hamiltonian (2) is equivalent to the
standard SK Hamiltonian.

In terms of the simplex spin $S_i^{\mu}$ ($1\leq \mu \leq p-1$), 
the  parameter $q$ may be
defined by
\begin{equation}
  q = \sqrt{\sum_{\mu,\nu}^{p-1}(q^{\mu\nu})^2},
  \label{eqn:sgorder}
\end{equation}
where $q^{\mu\nu}$ denotes an overlap tensor between two replicas 1 and 2, 
\begin{equation}
  q^{\mu\nu}=\frac{1}{N}\sum_{i=1}^{N}S_{i,1}^{\mu}S_{i,2}^{\nu}.
\end{equation}
The Binder parameter is then given by 
\begin{equation}
 g(T,N) = 
\frac{(p-1)^2}{2}\left(1+\frac{2}{(p-1)^2}-\frac{[\langle q^4\rangle]}
{[\langle q^2\rangle]^2}\right), 
\end{equation} 
where $\langle\cdots\rangle$ denotes the thermal average and $[\cdots]$
denotes the average over the quenched randomness $\{J_{ij}\}$.
The Binder parameter is normalized  so as 
to vanish above the transition temperature 
$T_{\rm g}$ in the thermodynamic
limit. 
Recall that  at $T>T_{\rm g}$ each component $q_{\mu\nu}$ should behave as an
independent Gaussian variable. 
Below $T_{\rm g}$, 
$g$ is normalized to
give unity in the thermodynamic limit
for the nondegenerate ordered state where  
$P(q)$ has only trivial
peak at $q=q_{{\rm EA}}$.  Of course, 
this is not the case for the SG models showing the RSB including the present
mean-field SG  models, for which $g$  
takes
nontrivial values different from unity
even in the thermodynamic limit.
Hence, at least in the case where a continuous phase transition
occurs into
the trivial ordered state,
$g$ for various finite sizes is expected to 
cross at $T=T_{\rm g}$. Indeed, this 
aspect has widely been 
used for locating the transition temperature from the numerical data for
finite systems. 

\section{Monte Carlo Results}
We perform MC simulations based on a version of the
extended ensemble method, called the exchange method\cite{eMC}. 
As in other SG models, an extremely slow relaxation becomes a
serious problem of MC simulations in
the present mean-field SG models.  Such 
difficulty could partly be overcome by using the exchange method,
which has turned out to be quite efficient in thermalizing  
various hardly relaxing systems.
The method enables us to study larger sizes and/or lower temperatures than
those attained previously. 
Our MC simulations have been performed up to $N=512$ at $T/J=0.25$ for the SK 
model, and $N=256$ at $T/J=0.4$ for the mean-field $p=3$ Potts-glass model,
where $T_{\rm g}/J=1$ in both models. 
Sample averages are taken over  $200-1792$ 
independent bond realizations depending on the size $N$. 
We note that the minimum temperature studied here are considerably lower
than the previous ones; {\it e.g.\/}, $N=512$ at $T/J=0.75$\cite{BhattYoung} 
for the SK model and $N=120$ at $T/J=0.98$ for the mean-field
$p=3$ Potts-glass model\cite{DJK}. 

\begin{figure} 
 \begin{center} 
    \leavevmode
    \epsfxsize=\figwidth
    \epsffile{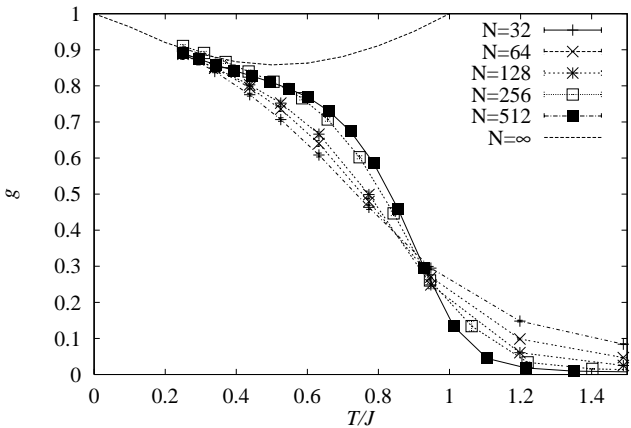}
 \end{center}
\caption{Temperature and size dependence of the Binder parameter of
 the SK model. The bulk transition temperature is located at $T/J=1$.
The broken line represents the RSB solution derived by
 solving the Parisi equation.}
\label{fig:BinderSK}
\end{figure}

\begin{figure}
    \epsfxsize=\figwidth
    \epsffile{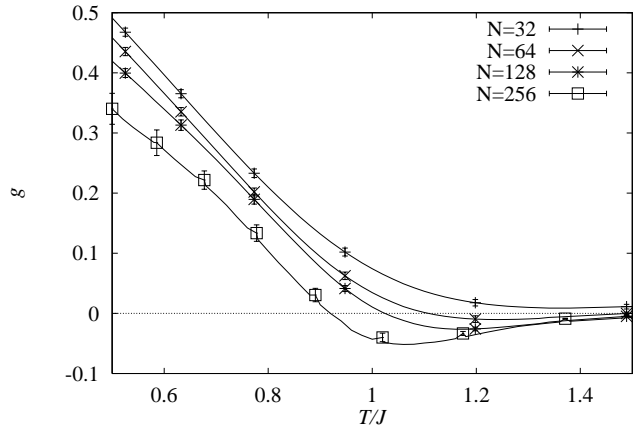}
  \caption{Temperature and size dependence of the Binder parameter of the
 mean-field $p=3$ Potts glass. The bulk 
transition temperature is located at $T/J=1$.}
  \label{fig:BinderPG}
\end{figure}

The temperature and size dependence of the calculated 
Binder parameter $g$ is shown 
in Figs.~\ref{fig:BinderSK} and~\ref{fig:BinderPG} for the SK and 
the $p=3$
Potts-glass
models, respectively. 
As is evident from these figures,  the Binder parameters 
of the two mean-field models
show considerably different behaviors from each other.

In the SK model, as shown in Fig.~\ref{fig:BinderSK},
a clear crossing  of $g$ is observed at $T=T_{\rm g}$,
which looks
similar to the ones seen in the standard continuous transitions.
In fact, the behavior of $g$ found here  also resembles the
ones observed in the 
short-range Ising SG models
in 3D\cite{KY,Marinari98} and  in 4D\cite{Marinari99},
though the crossing tendency is less pronounced in 3D than in 4D.
As mentioned, $g$ of the SK model takes a nontrivial value
below $T_{\rm g}$ even in the thermodynamic limit due to its RSB.
We show  in Fig.~\ref{fig:BinderSK} 
the behavior of $g(T,\infty)$ evaluated in the replica 
formalism 
by numerically solving the Parisi equation [1(a)].  
Note that, as the temperature approaches $T_{\rm g}$ from below, 
the limiting value $g(T_{\rm g}^-, \infty )$ goes to unity as in the
case of ordinary continuous phase transitions. 
Hence, with increasing $N$,
$g(T_{\rm g}^-, N)$ just below $T_{\rm g}$ is expected
to approach unity {\it from below\/} while $g(T_{\rm g}^+, N)$ 
just above $T_{\rm g}$ approaches zero {\it from above\/}, 
which entails a
crossing of $g$ at $T=T_{\rm g}$. 
With lowering the temperature, $g(T, \infty)$ 
first decreases, reaching a minimum around $T/J=0.5$, and increases again
tending to unity at $T=0$. Here note that, for any model with nondegenerate
ground state, $P(q)$ becomes trivial at $T=0$
irrespective of the occurrence of RSB, and $g$ tends to unity. 
As can be seen in Fig.~\ref{fig:BinderSK}, 
the present MC results for finite $N$ gradually
approach the $g(T, \infty)$ curve of an infinite system. 

In the mean-field $p=3$ Potts glass, 
as shown in Fig.~\ref{fig:BinderPG}, 
no crossing of $g$ is observed at $T=T_{\rm g}
$\cite{Peters}, at least of
the type as observed in the SK model.
Instead, unlike the case of the SK model, 
 a shallow {\it negative\/} 
dip develops above $T_{\rm g}$ for larger $N$ which 
becomes deeper as
the system gets larger.
Although the existence of 
a negative dip was not reported in the previous numerical works\cite{Peters}, 
we note that  a negative dip appears only for larger $N$ which
accounts for the absence of a negative dip in the previous data.
Perhaps, on looking at Fig.~\ref{fig:BinderPG}, one would 
hardly imagine that there occurs a continuous phase transition at
$T/J=1$: Nevertheless, the occurrence of a continuous transition at $T/J=1$
is an exactly established property of the model.
We also note that, 
while the appearance of a growing negative dip in the Binder parameter
is often related to the occurrence of a first-order
transition\cite{Vollmayr}, 
this is not always the case: Here, the transition is
established to be continuous. 

It might be instructive to examine here the behavior of $g$
in the thermodynamic limit.
As the temperature approaches  $T_{\rm g}$ 
from below,  $g(T_{\rm g}^-,\infty)$
tends to a {\it negative\/} 
value,  $-1$ in the present case.
Such a negative value of $g(T_{\rm g}^-,\infty)$
is in sharp contrast to the system showing the 
full RSB where $g(T_{\rm g}^-,\infty)=1$.
Indeed, this negativity is closely related to
the occurrence of  the one-step RSB in the model\cite{comment1}.

Then, one expects that
the negative dip of $g(T,N)$ observed in Fig.~\ref{fig:BinderPG}
further deepens with increasing $N$, and eventually approaches $-1$
{\it from above\/} at $T=T_{\rm g}^-$,
in sharp contrast to the SK case where $g(T_{\rm g}^-,N)$ approaches 1
{\it from below\/}. 
Therefore,  the crossing of $g$ in the $g>0$ region as observed 
in the SK model hardly occurs in the $p=3$ Potts-glass model.
Rather, if one considers the fact that 
$g(T,N)$  above $T_{\rm g}$ is negative for moderately large $N$ 
approaching zero {\it from below\/}, the crossing
of $g$ is expected to occur  {\it in the  $g<0$  region\/}, 
not in the $g>0$ region
as in the case
of the SK model. The data of Fig.~\ref{fig:BinderPG} are certainly
consistent with such a behavior. Anyway, our present result of the
mean-field $p=3$ Potts glass has revealed that
the data of the Binder parameter has to be interpreted with special care
particularly when the ordered state has  one-step RSB features.

Next, we study the so-called Guerra parameter which was originally introduced 
to detect the RSB transition\cite{Marinari98a},
\begin{equation}
 G(T,N) = \frac{[\langle q^2\rangle^2]-[\langle q^2\rangle]^2}
{[\langle q^4\rangle]-[\langle q^2\rangle]^2}.
\label{eqn:Guerra}
\end{equation}
Since the numerator represents a sample-to-sample fluctuation of the overlap, 
non-vanishing of $G$ means a lack of self-averaging
so long as the denominator remains nonzero. 
In the mean-field SG models  studied here, their 
RSB indeed gives rise to the lack of
self-averaging, {\it i.e.,\/} the occurrence of a non-trivial probability
distribution of the overlap over quenched
disorder. It has been rigorously proven,
without using the replica trick, that in the SG phase of 
the SK model the $G$ parameter in the thermodynamic limit is equal to 1/3
independent of the temperature\cite{Guerra}.   
Meanwhile, it has been pointed out in Ref.~\cite{Bokil} that, 
even when $P(q)$ is trivial and the ordered state is self-averaging,
the  $G$ parameter can still take a nonzero value due to the  
possible vanishing
of the denominator, leading to a crossing at
$T_{\rm g}$\cite{Bokil}. Hence, the crossing of $G$ does not necessarily
mean the lack of the self-averaging, although
it can still be used as an indicator  of a phase transition.
As an indicator of the non-self-averageness in the ordered state,
one may use the $A$ parameter defined by\cite{reply}
\begin{equation}
 A(T,N) =  \frac{[\langle q^2\rangle^2]-[\langle q^2\rangle]^2}
  {[\langle q^2\rangle]^2}. 
\label{eqn:NSA}
\end{equation}

\begin{figure}
 \epsfxsize=\figwidth
 \epsffile{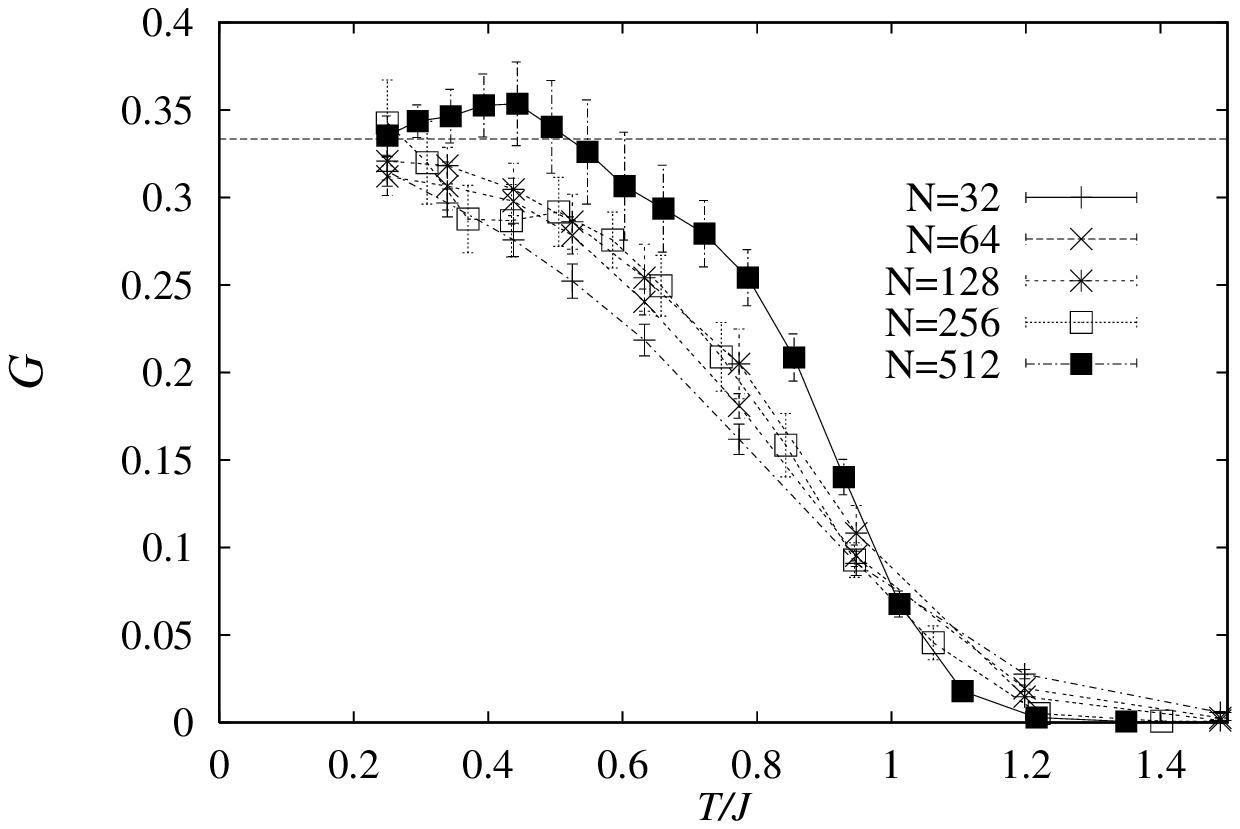}
 \caption{Temperature and size dependence of the $G$ parameter,
defined by Eq.~(\protect{\ref{eqn:Guerra}}), of
 the SK model. The bulk transition temperature is located at $T/J=1$.
 The broken straight line represents the line $G=1/3$.}
 \label{fig:G-sk}

 \epsfxsize=\figwidth
 \epsffile{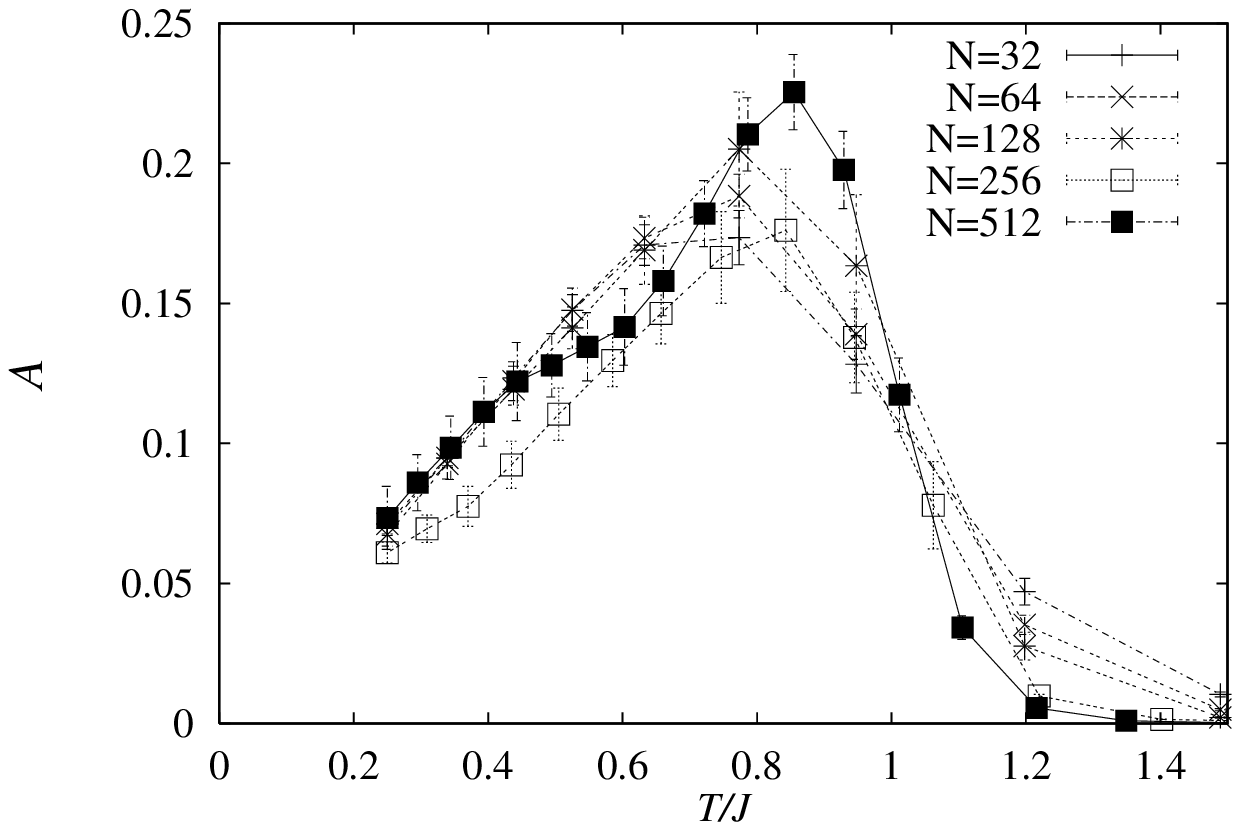}
 \caption{Temperature and size dependence of the $A$ parameter,
defined by Eq.~(\protect{\ref{eqn:NSA}}), of
 the SK model. The bulk transition temperature is located at $T/J=1$.}
 \label{fig:A-sk}
\end{figure}

We calculate these two parameters, $G$ and $A$, 
both for the SK and the mean-field $p=3$
Potts-glass models. 
The temperature and size
dependence of the $G$ and $A$ parameters 
of the SK model is shown in Figs.~\ref{fig:G-sk} and~\ref{fig:A-sk}, 
respectively. 
Although the error bars are still
large, 
both $G$ and $A$ show a clear crossing at $T_{\rm g}$, 
remaining positive at any temperature.
As expected, with increasing $N$,
the  $G$ parameter approaches $1/3$ independent of $T$
below $T_{\rm g}$. By contrast, the  $A$ parameter for various sizes 
merge into a curve  below $T_{\rm g}$, 
which clearly stays nonzero indicating the 
non-self-averageness of the ordered state. 
Here it should be noticed that,
just at the transition point, 
the non-self-averageness is expected to occur
in any random system, 
even including the ones without showing the RSB in the ordered
state\cite{Wiseman,Aharony}. 
Hence, in the type
of random systems which do not show the RSB in the ordered state,
$A(T, \infty)$ stays nonzero only just at $T=T_{\rm g}$
and vanishes on both sides of $T_{\rm g}$. By contrast, in the present
SK model, $A(T, \infty)$ should stay finite even below $T_{\rm g}$ due
to its RSB, which explains the observed merging behavior seen in
Fig.~\ref{fig:A-sk} at $T<T_{\rm g}$. As can be seen from
Fig.~\ref{fig:A-sk}, on further lowering the temperature toward $T=0$,
$A(T,N)$ tends to vanish in contrast to the behavior of $G(T,N)$.
This aspect is consistent with the fact that at $T=0$ the overlap
distribution becomes trivial and the self-averageness is recovered
irrespective of the occurrence of RSB. 

\begin{figure} 
  \epsfxsize=\figwidth
  \epsffile{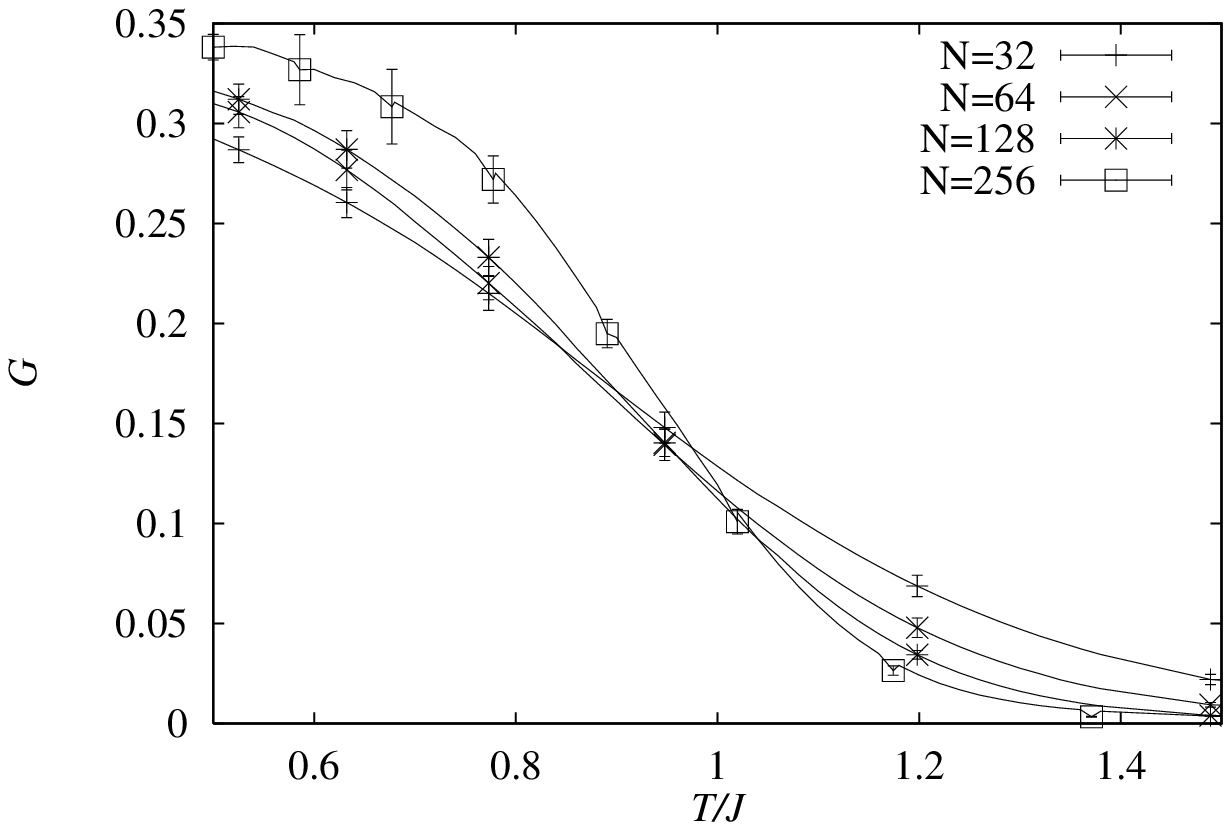}
  \caption{Temperature and size dependence of $G$ parameter,
defined by Eq.~(\protect{\ref{eqn:Guerra}}), of the
  mean-field $p=3$ Potts glass. 
The bulk transition temperature is located at $T/J=1$.}
  \label{fig:G-pg}
\end{figure}

\begin{figure} 
  \epsfxsize=\figwidth
  \epsffile{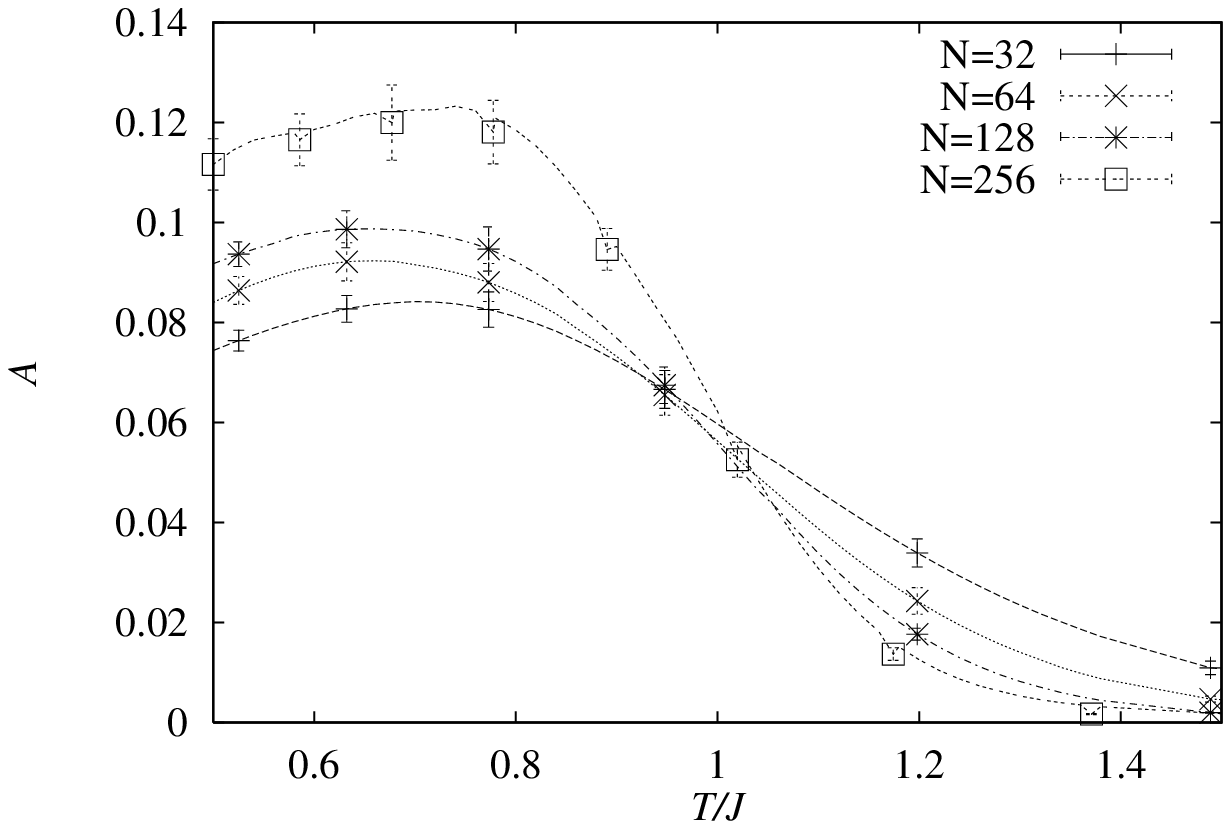}
  \caption{Temperature and size dependence of $A$ parameter, defined by
Eq.~(\protect{\ref{eqn:NSA}}), of the
  mean-field $p=3$ Potts glass. 
The bulk transition temperature is located at $T/J=1$.}
  \label{fig:A-pg}
\end{figure}

The $G$ and $A$ of the mean-field $p=3$ Potts glass  are presented in
Figs.~\ref{fig:G-pg} and \ref{fig:A-pg}, respectively.   
Unlike the case of the Binder parameter $g$ shown in Fig.~\ref{fig:BinderPG},
the $G$ and $A$ parameters  remain positive at any $T$
and show a clear crossing at $T=T_{\rm g}$:
They behave more like the Binder parameter of  standard systems, {\it e.g.\/},
like the one shown in Fig.~\ref{fig:BinderSK}. 
In fact, the behaviors of the $G$ and $A$ parameters shown in
Figs.~\ref{fig:G-pg} and  \ref{fig:A-pg}
are  similar to those of the SK model shown in Figs.~\ref{fig:G-sk} and
\ref{fig:A-sk}, suggesting that  
$G$ and $A$ are less sensitive to the kind of breaking pattern of replica
symmetry.
Hence, one could use 
the $G$ and $A$ parameters  to identify the SG transition
based on the standard crossing method even for  
systems showing a one-step RSB.

\begin{figure}
    \epsfxsize=\figwidth
    \epsffile{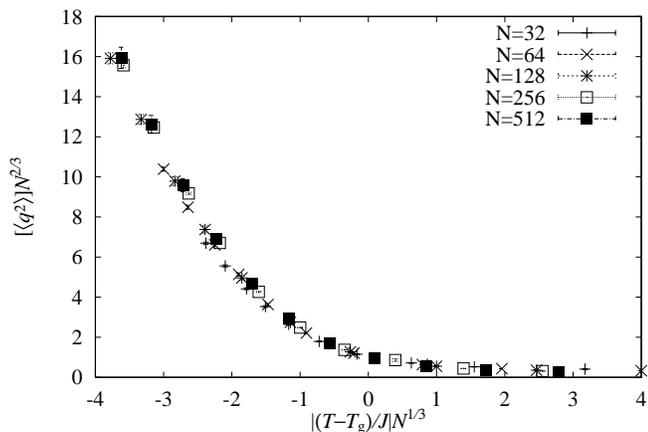}
\caption{
Finite-size scaling plot of the squared order parameter of the SK model with
 the scaling form Eq.~(\protect{\ref{eqn:fss}}) with $T_g/J=1$.
}
\label{fig:scal-sk}
\end{figure}

\begin{figure}
    \epsfxsize=\figwidth
    \epsffile{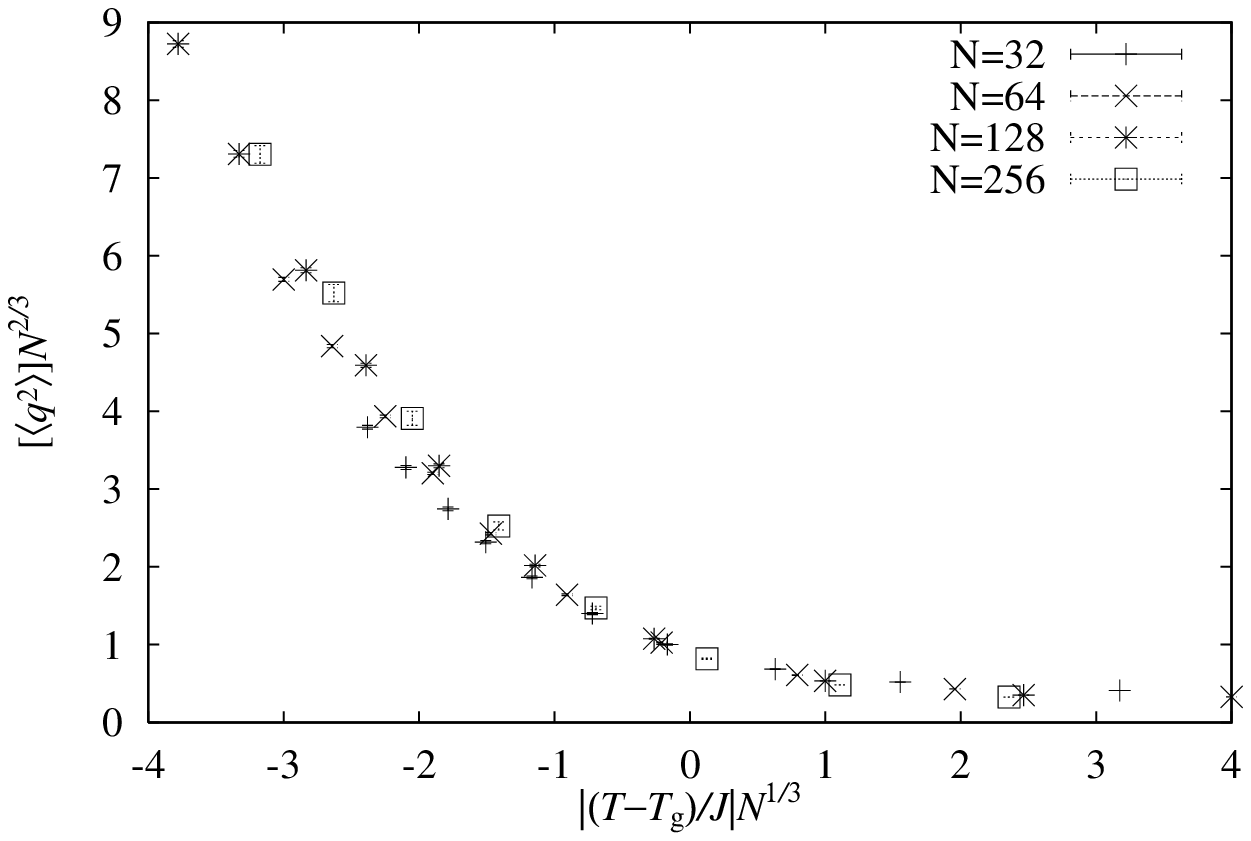}
 \caption{
Finite-size scaling plot of the squared order parameter of the  $p=3$ 
mean-field Potts glass with  the scaling form Eq.~(\ref{eqn:fss})
with $T_g/J=1$.
}
\label{fig:scal-pg}
\end{figure}

Once the transition temperature is established, the next  task 
would be to determine critical exponents. 
Here we wish to examine a finite-size scaling 
hypothesis concerning the SG order
parameter for the present mean-field models. Similar analysis has 
widely been used for extracting the critical exponents from the
numerical data.  According to Ref.~\cite{Botet}, 
finite-size scaling of the mean-field models can be derived by assuming
that the ``coherence number''  behaves as $\xi^{d_u}$ where 
$d_u$ is the upper critical dimension of the corresponding short-range model, 
while the ``coherence length'' $\xi$ diverges at $T=T_{\rm g}$ 
with the correlation-length 
exponent at the upper critical dimension $\nu_{\rm MF}$,
$\xi\sim |T-T_{\rm g}|^{-\nu_{\rm MF}}$.
Then, the squared order parameter can be written as 
\begin{eqnarray}
[\langle q^{2}\rangle]  
& \sim |T-T_{\rm g}|^{2\beta _{\rm MF}} f(N|T-T_{\rm g}|^{d_u\nu_{\rm MF}}), 
\nonumber \\ 
& \sim N^{-2\beta _{\rm MF}/d_u\nu_{\rm MF}} 
f'(N|T-T_{\rm g}|^{d_u\nu_{\rm MF}}),
\end{eqnarray}
where $\beta _{\rm MF}=1$ is the mean-field order-parameter exponent whereas
$f$ and $f'$ are the scaling functions.
Noting the fact that
the upper critical dimension of the SG models is $d_u=6$
and the correlation-length exponent at $d=d_u=6$ is equal to
$\nu_{\rm MF}$=1/2, it follows
\begin{eqnarray}
[\langle q^{2}\rangle]  
\sim N^{-2/3}f''(|T-T_{\rm g}|N^{1/3}).
\label{eqn:fss}
\end{eqnarray}
The resulting finite-size 
scaling plots are shown in Figs.~\ref{fig:scal-sk} and
\ref{fig:scal-pg} for the SK and the $p=3$ Potts-glass models,
respectively. 
In both models, the scaling of the form (\ref{eqn:fss}) turns out to
work fairly well both {\it below} and {\it above} $T_{\rm g}$ as far as
the temperature is sufficiently close to $T_{\rm g}$. We note that a
similar finite-size-scaling analysis has already been reported for the SK
model  just at $T_{\rm g}$\cite{Parisi93} and for the $p=3$ Potts-glass
model above $T_{\rm g}$ \cite{DJK}.  
In particular, the scaling turns out to be reasonably
good even for the $p=3$ Potts glass where the Binder parameter does 
not exhibit a clear crossing in the range of sizes studied.
This implies that 
the standard finite-size scaling analysis of the order parameter
could still be useful even in  RSB systems including the one-step RSB systems. 


\section{Discussion and Remarks}
In this section, with our present results for the mean-field models in mind, 
we wish to
comment on the possible RSB in some {\it short-range\/} SG models. 

As mentioned, one-step RSB-like features were recently observed 
in the chiral-glass state of the 3D short-range Heisenberg SG\cite{HK}.
There, the Binder parameter for the chirality, the order parameter of
the chiral-glass transition, 
did not  cross in the  $g>0$ region and
developed a negative dip which deepened with the system size.
Instead, a crossing of $g$ was observed in the $g<0$ region close
to the negative dip (see Fig.~1 of Ref.~\cite{HK} ). 
Meanwhile, the $G$ parameter always remained
positive and showed a clear crossing at $T=T_{\rm g}$ (see Fig.~3 of 
Ref.~\cite{HK} ). 
All these features are similar to the ones observed here in the
mean-field $p=3$ Potts glass, suggesting that the chiral-glass 
state of the 3D Heisenberg SG has a one-step RSB-like
character\cite{comment2}. 

Other obvious interest is the nature of the possible 
phase transition of the  short-range three-state ($p=3$) Potts-glass 
model in 3D.
It is widely believed that there is no finite-temperature phase
transition in 3D $p=3$ Potts glass which
were investigated by MC
simulations\cite{Scheucher90,Scheucher92a,Scheucher92b} and other
numerical methods\cite{Singh91}.  
In particular,
MC results of Refs.~\cite{Scheucher90,Scheucher92b} revealed that the
Binder parameter decreased 
monotonically with system size  without showing a crossing, which was
taken as an evidence of the absence of a finite-temperature transition.
However, the behavior of $g$  observed in
Refs.~\cite{Scheucher90,Scheucher92b} was not dissimilar to the one 
observed here in the mean-field $p=3$ Potts glass, and we feel that 
the possibility
of the occurrence of a one-step RSB-like transition at finite $T_{\rm g}$
still cannot be ruled out.

Recently, 
short-range $p$-spin glass models whose mean-field versions have been
known to show the one-step RSB were studied by MC
simulations\cite{Campellone,Parisi99}. 
For example,  
according to the calculation of Ref.~\cite{Parisi99} for the 4D $3$-spin
model,  the Binder parameter did not exhibit a crossing of the standard
type, while the $G$ and $A$ parameters  showed a clear crossing at
$T=T_{\rm g}>0$, 
strongly suggesting the
occurrence of a finite-temperature transition. 
Thus, from out present study, the possible occurrence of a one-step RSB
transition at $T=T_{\rm g}>0$ is suspected.
Meanwhile, a closer inspection reveals that
a negative dip observed in $g$
becomes shallower with increasing system size\cite{Parisi99}, 
in contrast to the case of
the mean-field $p=3$ Potts glass studied here. 
Further studies seems to be required to clarify the nature of
the RSB in the short-range $p$-spin glass.


In conclusion, we have investigated by MC simulations
the finite-size effects of the two mean-field SG
models whose replica-symmetry-breaking properties in the 
thermodynamic limit are well established. 
In the mean-field Ising spin glass (the SK model), the Binder parameter
$g$ of various sizes always remains positive and crosses at 
$T=T_{\rm g}$, while
in the mean-field three-state Potts glass, it develops a negative dip which
deepens as the system size increases, without a crossing in the
$g>0$ region as observed in the SK model. Instead, a crossing of $g$ occurs
in the $g<0$ region near the negative dip. 
Such difference in the behaviors of $g$ reflects the different types of 
associated RSB of the two models,
{\it i.e.\/}, full versus one-step RSB.
By contrast, the Guerra parameter $G$ as well as the non-self-averaging
parameter $A$ always remain positive and show a crossing of the standard
type at $T=T_{\rm g}$ for both the SK and 
$p=3$ Potts-glass models. 
We have also discussed  implications of the present results
to the possible interpretation of the numerical results
for some short-range SG models.

\acknowledgments
The authors would like to thank H.~Takayama and H.~Yoshino for valuable
discussions. 
One of the present authors (KH) was supported by Grant-in-Aid for the
Encouragement of Young  Scientists from the Ministry of Education, 
Science, Sports and Culture of  Japan (No.~11740220). 
Numerical simulation was performed on several workstations at ISSP,
University of Tokyo.


\begin{thebibliography}{99}
\bibitem{comment1}
In the mean-filed $p$ state Potts-glass model with $p<4$, 
$g(T_{\rm g}^-)$ is given by  $g(T_{\rm g}^-)=\frac{-p^3+4p^2-7p+10}{2(4-p)}$. 
Hence, $g(T_{\rm g}^-)$ takes a negative value as long as $p>2.78$. 
On the other hand, 
in the random energy model corresponding to $p=\infty$ Potts-glass model, 
$g(T_{\rm g}^-,\infty)= -\infty$.
Of course,
it is still possible that $g(T_{\rm g}^-)$   takes a positive value 
in some continuous one-step RSB transitions.
\end{thebibliography}
\end{document}